\documentclass[sigconf]{acmart}
\usepackage{booktabs} % For formal tables
\usepackage{spreadtab}
\usepackage{amsmath}
\usepackage{float}
\usepackage{todonotes}
\usepackage{graphicx}
\usepackage{caption}
\usepackage{color}
\usepackage{subcaption}
\DeclareCaptionSubType*[Alph]{table}
\DeclareCaptionLabelFormat{mystyle}{Table~\bothIfFirst{#1}{ ̃}#2}
\copyrightyear{2018} 
\acmYear{2018} 
\setcopyright{acmcopyright}
\acmConference[SIGIR '18]{The 41st International ACM SIGIR Conference on Research & Development in Information Retrieval}{July 8--12, 2018}{Ann Arbor, MI, USA}
\acmPrice{15.00}
\acmDOI{10.1145/3209978.3210107}
\acmISBN{978-1-4503-5657-2/18/07}

\begin{document}
\title{Consistency and Variation in Kernel Neural Ranking Model}
% \author{Mary Arpita Pyreddy\footnotemark[1], Varshini Ramaseshan\footnotemark[1], Narendra Nath Joshi}
% \authornote{The first three authors contributed equally}
% \affiliation{%
%    \institution{Carnegie Mellon University}
%    \streetaddress{5000 Forbes Ave}
%    \city{Pittsburgh}
%    \state{Pennsylvania}
%    \postcode{15213}
%  }
% \email{mpyreddy, vramases, nnj@cs.cmu.edu}

% \author{Zhuyun Dai, \\Chenyan Xiong, \\Jamie Callan} 
% \affiliation{%
%    \institution{Carnegie Mellon University}
%    \streetaddress{5000 Forbes Ave}
%    \city{Pittsburgh}
%    \state{Pennsylvania}
%    \postcode{15213}
%  }
%    \email{zhuyund, cx, callan@cs.cmu.edu}

\author{Mary Arpita Pyreddy}
\authornote{The first three authors contributed equally}
\affiliation{%
   \institution{Carnegie Mellon University}
%    \streetaddress{5000 Forbes Ave}
%    \city{Pittsburgh}
%    \state{Pennsylvania}
%    \postcode{15213}
   }
   \email{mpyreddy@cs.cmu.edu}

\author{Varshini Ramaseshan\footnotemark[1]} 
\affiliation{%
   \institution{Carnegie Mellon University}
%    \streetaddress{5000 Forbes Ave}
%    \city{Pittsburgh}
%    \state{Pennsylvania}
%    \postcode{15213}
   }
   \email{vramases@cs.cmu.edu}

\author{Narendra Nath Joshi\footnotemark[1]}     
\affiliation{%
   \institution{Carnegie Mellon University}
%    \streetaddress{5000 Forbes Ave}
%    \city{Pittsburgh}
%    \state{Pennsylvania}
%    \postcode{15213}
   }
\email{nnj@cs.cmu.edu}

\author{Zhuyun Dai}
\affiliation{%
   \institution{Carnegie Mellon University}
%    \streetaddress{5000 Forbes Ave}
%    \city{Pittsburgh}
%    \state{Pennsylvania}
%    \postcode{15213}
   }
\email{zhuyund@cs.cmu.edu}

\author{Chenyan Xiong}
\affiliation{%
   \institution{Carnegie Mellon University}
%    \streetaddress{5000 Forbes Ave}
%    \city{Pittsburgh}
%    \state{Pennsylvania}
%    \postcode{15213}
   }
\email{cx@cs.cmu.edu}

\author{Jamie Callan}
\affiliation{%
   \institution{Carnegie Mellon University}
%    \streetaddress{5000 Forbes Ave}
%    \city{Pittsburgh}
%    \state{Pennsylvania}
%    \postcode{15213}
   }
\email{callan@cs.cmu.edu}

\author{Zhiyuan Liu}
\affiliation{%
   \institution{Tsinghua University}
%    \city{Beijing}
%    \country{China}
 }
    \email{liuzy@tsinghua.edu.cn}

\renewcommand{\shortauthors}{Pyreddy et al.}

\begin{abstract}
This paper studies the consistency of the kernel-based neural ranking model (\texttt{K-NRM}), a recent state-of-the-art neural IR model, which 
is important for reproducible research and deployment in the industry.  We find that \texttt{K-NRM} has low variance on relevance-based metrics across experimental trials.  In spite of this low variance in overall performance, different trials produce different document rankings for individual queries. The main source of variance in our experiments was found to be different latent matching patterns captured by \texttt{K-NRM}. In the IR-customized word embeddings learned by \texttt{K-NRM}, the query-document word pairs follow two different matching patterns that are equally effective, but align word pairs differently in the embedding space.
The different latent matching patterns enable a simple yet effective approach to construct ensemble rankers, which improve \texttt{K-NRM}'s effectiveness and generalization abilities.
\end{abstract}


% <concept_id>10002951.10003317</concept_id>
% <concept_desc>Information systems~Information retrieval</concept_desc>
% <concept_id>10002951.10003317.10003338</concept_id>
% <concept_desc>Information systems~Retrieval models and ranking</concept_desc>
% <concept_id>10002951.10003317.10003325</concept_id>
% <concept_desc>Information systems~Information retrieval query processing</concept_desc>
% <concept_significance>300</concept_significance>

% \ccsdesc[500]{Information systems~Information retrieval}
% \ccsdesc[500]{Information systems~Retrieval models and ranking}
% \ccsdesc[300]{Information systems~Information retrieval query processing}


\maketitle

\section{Introduction}
Neural IR models have received much attention due to their continuous text representations, soft-matching of terms, and sophisticated non-linear models.  However, the non-convexity and stochastic training of neural IR models raises questions about their consistency compared to heuristic and learning-to-rank models that use discrete representations and simpler methods of combining evidence.  Consistent behavior under slightly different conditions is essential to reproducible research and deployment in industry.

This paper studies the stability of \texttt{K-NRM}, a recent state-of-the-art neural ranking model~\cite{xiong2017end}. 
\texttt{K-NRM} learns the word embeddings and ranking model from relevance signals. Its effectiveness is due to word embeddings tailored for search tasks and kernels that group matches into bins of different quality.  Its parameter space is large, the solution space is non-convex, and training is stochastic.

To better understand its stability, we compare the behavior of multiple trained models under similar conditions. We find that although \texttt{K-NRM} produces similar accuracy across different trials, it also produces rather different document rankings for individual queries. 

%An analysis of the learning-to-rank weights for \texttt{K-NRM} kernel scores (soft-match features) revealed that the learned weights from different trials match one of two patterns.
%The word embeddings reflect these patterns. Trials whose kernel weights fall into the same pattern produce similar word embeddings.
%Interestingly, the two patterns are equally effective.

Analysis of weights learned for \texttt{K-NRM} kernel scores (soft-match features) revealed that weights from different trials match one of two patterns.
The word embeddings reflect these patterns. Trials whose kernel weights have the same pattern have similar word embeddings.
Interestingly, the two patterns are equally effective.

The difference in the ranking patterns from different \texttt{K-NRM} trials makes them a good fit for ensembles.  Aggregating scores from different trials enables an ensemble to
promote documents that multiple trials agree are most likely to be relevant.
Experimental results show that simple \texttt{K-NRM} ensembles significantly boost its ranking accuracy and improve its generalization ability. 
\section{Related Work}\label{section:related-work}
%Recent neural IR methods can be categorized as \emph{representation-based} and 
%\emph{interaction-based}~\cite{guo2016deep}.
%\textit{Representation-based} models focus on learning good distributed representations of the query and document and matching them in the representation space~\cite{huang2013learning,shen2014latent}. \textit{Interaction-based models} focus on building local interactions between the query and document words and then use neural networks to learn complex matching patterns~\cite{ pang2016,guo2016deep,xiong2017end}.

Recent neural IR methods can be categorized as \emph{representation-based} and 
\emph{interaction-based}~\cite{guo2016deep}.
\textit{Representation-based} models use distributed representations of the query and document that are matched in the representation space~\cite{huang2013learning,shen2014latent}. \textit{Interaction-based models} use local interactions between the query and document words and neural networks that learn matching patterns~\cite{guo2016deep,xiong2017end}.

%=====cx the language needs polish in this paragraph.
% \texttt{K-NRM} \cite{xiong2017end} is an interaction-based model which uses a kernel pooling technique to summarize word-word interactions. The model builds word-word similarity matrix from word embeddings, and uses kernel pooling to `count' the soft matches at multiple similarity levels using Gaussian kernels with mean value $\mu$ ranging from the cosine range $[-1, 1]$. A linear learning-to-rank layer is used to combine the kernel features. The whole model is end-to-end trainable. When trained from a search log, \texttt{K-NRM} outperforms neural IR methods and feature-based learning-to-rank methods. 

\texttt{K-NRM} \cite{xiong2017end} is an interaction-based model that uses kernel pooling to summarize word-word interactions. It builds a word-word similarity matrix from word embeddings, and uses kernel pooling to `count' the soft matches at multiple similarity levels using Gaussian kernels. A linear learning-to-rank layer combines the kernel features. The whole model is end-to-end trainable. When trained from a search log, \texttt{K-NRM} outperforms neural IR methods and feature-based learning-to-rank methods. 

% At the current stage, most Neural IR research focus on the effectiveness of the models.
% Meanwhile, the high variance of deep learning models raises concern about their consistency. Haber et al. \cite{haber2017stable} identify the causes for lack of stability and high variance as the dimensionality and non-convexity of the optimization problem. 
% A common method to reduce variance and improve generalization is to create an ensemble of the base models~\cite{krogh1995neural}. Krogh and Vedelsby~\cite{krogh1995neural} argue that a good ensemble is one where the individual networks are all accurate but disagree on individual examples. Ensemble of neural networks has been successfully applied to various tasks such as image classification~\cite{krizhevsky2012imagenet} and machine translation~\cite{sutskever2014sequence}.

Most neural IR research focuses on ranking accuracy.
However, the high variance of deep learning models causes concern about their consistency. Haber et al. \cite{haber2017stable} identify the causes for lack of stability and high variance as the dimensionality and non-convexity of the optimization problem. 
A common method to reduce variance and improve generalization is to create an ensemble of models~\cite{krogh1995neural}. Krogh and Vedelsby~\cite{krogh1995neural} argue that a good ensemble is one where the components are all accurate but disagree on individual examples. Ensembles of neural network have been applied successfully to tasks such as image classification~\cite{krizhevsky2012imagenet} and machine translation~\cite{sutskever2014sequence}.
\begin{table*}
\centering
\caption{Statistics from 50 \texttt{K-NRM} trials trained with random parameter initialization. 
% All trials were configured and evaluated in exactly the same procedure.
Minimum, Mean, and Maximum are the worst, average, and best performances. Standard Deviation is calculated on the corresponding 50 evaluation scores. }
\label{table:stability-runs}
\begin{tabular}{@{}l|c|c|c||c|c|c||c@{}}
\hline
 & \multicolumn{3}{c||}{\textbf{Testing-SAME}} & \multicolumn{3}{c||}{\textbf{Testing-DIFF}} & \textbf{Testing-RAW} \\ \hline
\multicolumn{1}{c|}{\textbf{Statistic}} & \multicolumn{1}{c|}{\textbf{NDCG@1}} & \multicolumn{1}{c|}{\textbf{NDCG@3}} & \multicolumn{1}{c||}{\textbf{NDCG@10}} & \multicolumn{1}{c|}{\textbf{NDCG@1}} & \multicolumn{1}{c|}{\textbf{NDCG@3}} & \multicolumn{1}{c||}{\textbf{NDCG@10}} & \multicolumn{1}{c}{\textbf{MRR}} \\ \hline
Minimum & 0.2531 & 0.3352 & 0.4221 & 0.2983 & 0.3234 & 0.4257 & 0.3430 \\
Mean & 0.2859 & 0.3495 & 0.4396 & 0.3242 & 0.3365 & 0.4378 & 0.3547 \\
Maximum & 0.3166 & 0.3744 & 0.4577 & 0.3484 & 0.3532 & 0.4496 & 0.3702 \\
Standard Deviation & 0.0130 & 0.0096 & 0.0104 & 0.0108 & 0.0076 & 0.0052 & 0.0067 \\ \hline
Reported in Xiong, et al.\cite{xiong2017end} & 0.2642 & 0.3210 & 0.4277 & 0.2984 & 0.3092 & 0.4201 & 0.3379 \\ \hline
\end{tabular}
\end{table*}
\begin{figure*}[!t]
\centering
  \begin{subfigure}[b]{0.30\linewidth}
    \includegraphics[width=\linewidth]{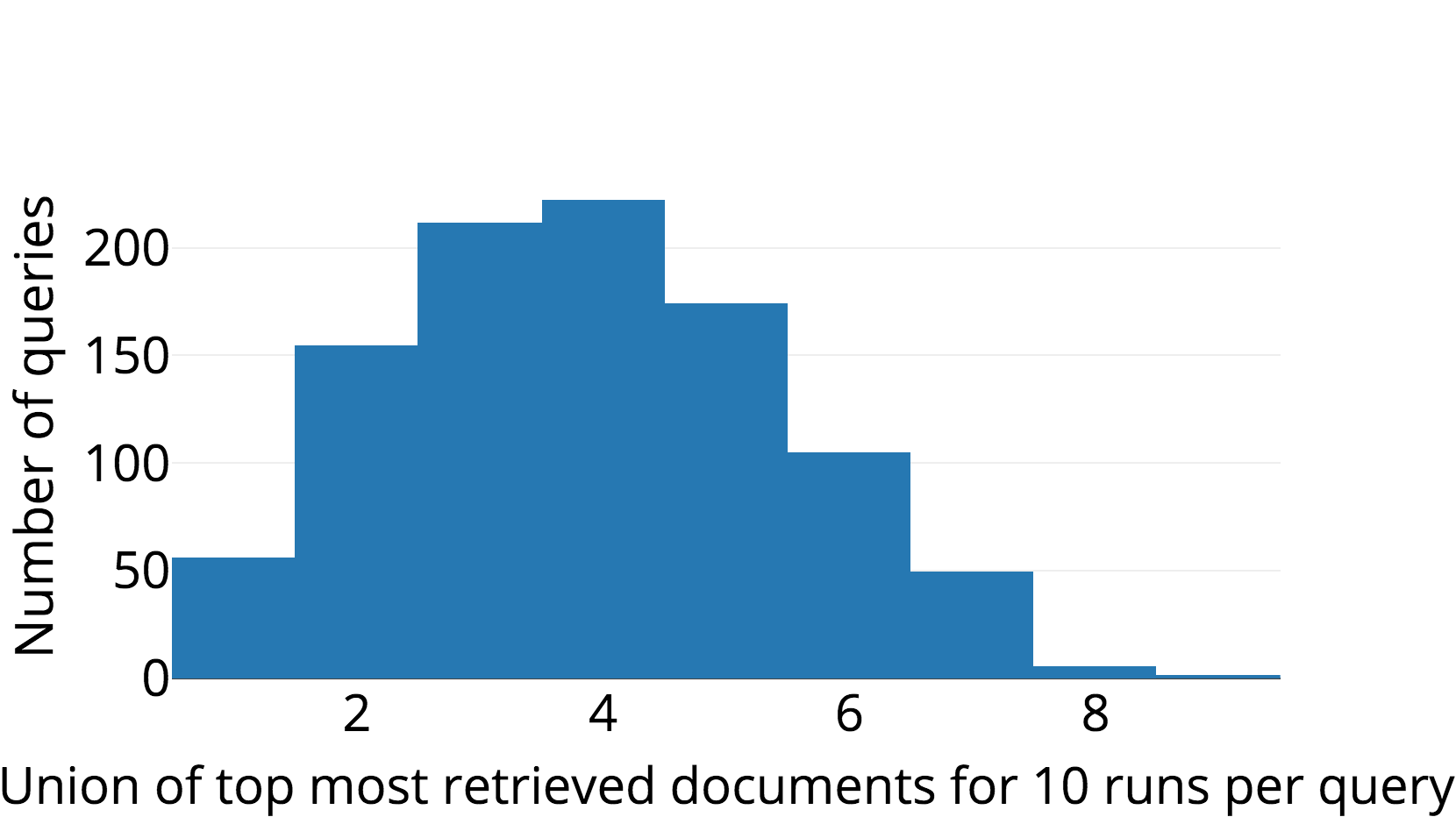}
    \caption{Agreement on Top 1}
    \label{fig:histogram1}
  \end{subfigure}
  \begin{subfigure}[b]{0.30\linewidth}
    \includegraphics[width=\linewidth]{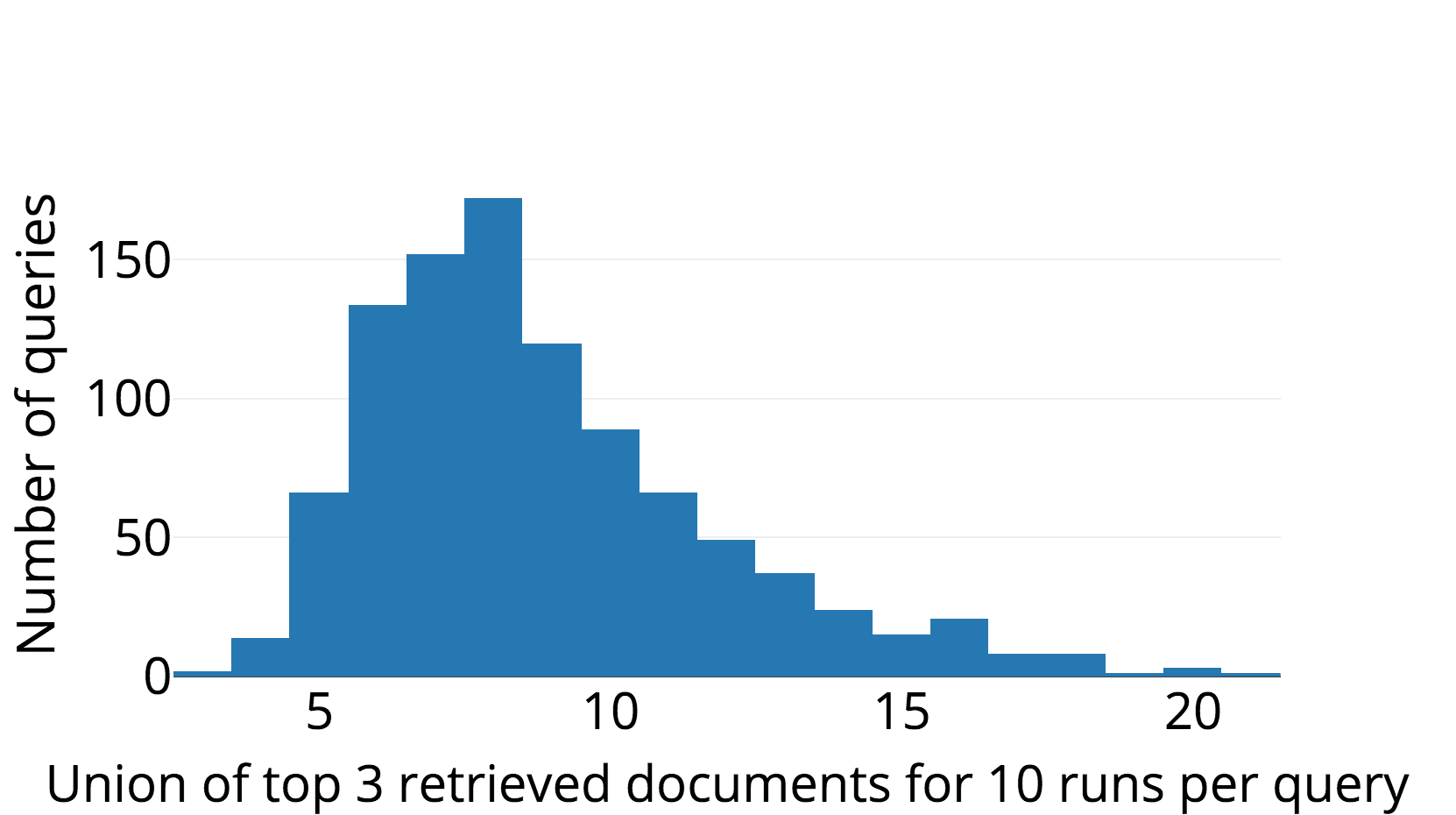}
%     \caption{Histogram of union of the top 3 documents for every query from the different runs}
 \caption{Agreement on Top 3}
    \label{fig:histogram3}
   \end{subfigure}
    \begin{subfigure}[b]{0.30\linewidth}
    \includegraphics[width=\linewidth]{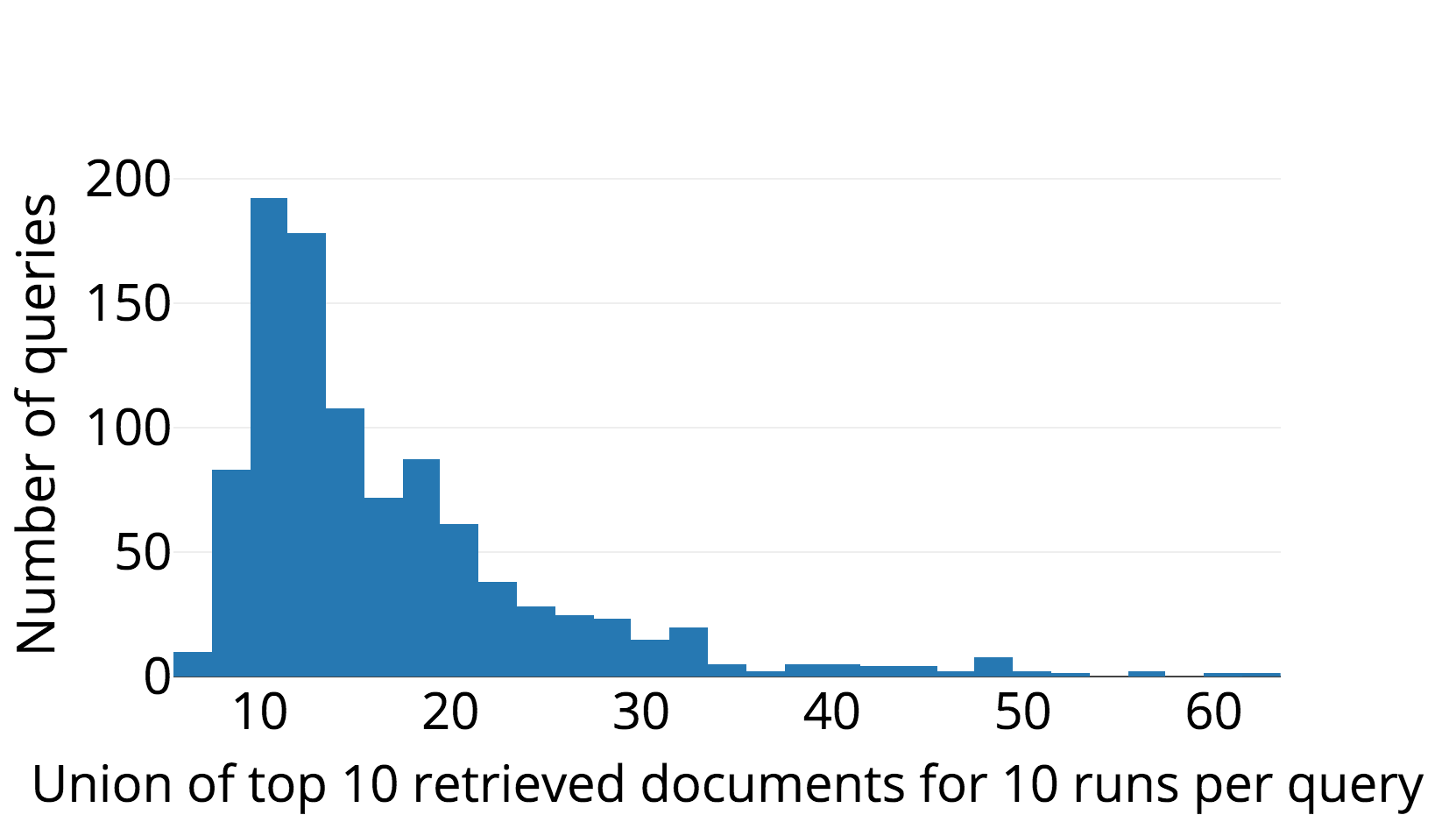}
%     \caption{Histogram of union of the top 10 documents from the different runs}
 \caption{Agreement on Top 10}
    \label{fig:histogram10}
  \end{subfigure}
 \caption{
 Query level ranking agreement. The X-axes are the number of distinct documents that appeared in the top K ranking results of 10 \texttt{K-NRM} trials. Larger X values indicate lower agreement among trials. 
 %Y-axes are the number of queries with rankings at the corresponding agreement level.
 \label{fig:agreement}
 }
\end{figure*}
\section{Experimental Setup} \label{section:exp-setup}

Our experiments followed the original \texttt{K-NRM} work \cite{xiong2017end} and used its open-source implementation\footnote{https://github.com/AdeDZY/K-NRM}. We used the same click log data from Sogou.com, a Chinese web search engine. The training set contained 95M queries, each with 12 candidate documents on average.  The testing set contained 1,000 queries, each with 30 candidate documents on average. Documents were represented by titles. Xiong, et al.~\cite{xiong2017end} built the vocabulary from queries and titles, but we built it from the queries, titles and URLs for better term coverage.

% \begin{table}
% \centering
% \caption{Statistics on the Sogou Query Log dataset \cite{xiong2017end} \cx{Do we need this table?} \zhuyun{No? Mention data size in the content.}}
% \label{table:stats}
% \begin{tabular}{@{}lll@{}}
% \toprule
%  & \textbf{Training} & \textbf{Testing} \\ 
%  \midrule
% \textbf{Number of queries} & 95,229 & 1000\\
% \textbf{Documents per query} & 12.17 & 30.50\\
% \textbf{Search sessions} & 31,201,876 & 4,103,230 \\
% \textbf{Vocabulary size} & 165,877 & 19,079 \\ 
% \bottomrule
% \end{tabular}
% \end{table}

\textbf{Training Labels}: The relevance labels for training were generated by the DCTR~\cite{chuklin2015click} click model from user clicks in the training sessions. 
DCTR uses the clickthrough rate for each query-document pair as the relevance score. 

\textbf{Testing Labels}: Following Xiong et al.~\cite{xiong2017end}, three testing conditions were used. \textbf{Testing-SAME} used DCTR to generate the testing labels while \textbf{Testing-DIFF} employed a more sophisticated model, TACM~\cite{liu2016tacm}. TACM takes into account both clicks and dwell times to generate testing labels. \textbf{Testing-RAW} treated the only clicked document in each single-clicked session as a relevant document, and used MRR (Mean Reciprocal Rank) as the metric. 
Testing-DIFF and Testing-RAW were considered more reliable than Testing-SAME because they are less subject to over-fitting. 

\textbf{Model Configuration}:  
We adopted the same default hyper-parameter configuration and 11 Gaussian kernels as in prior work \cite{xiong2017end}. The first kernel had $\mu=1, \sigma=10^{-3}$ to cover exact matches. The other 10 kernels were equally split in the cosine value range $[-1, 1]$:  $\mu_1=0.9,\mu_2=0.7,...,\mu_{10}=-0.9$; $\sigma$ was set to $0.1$. Word embeddings were initialized with a pretrained word2vec. The model used the Adam optimizer and was trained with batch size 16, learning rate = 0.001, and $\epsilon=1e-5$. The order of training data batches was fixed. An early stopping condition of 2 epochs 
was used in all experiments. 

The model was implemented using TensorFlow. All experiments were executed on a p2.xlarge AWS instance with 4 virtual CPUs and 1 NVIDIAGK210 GPU. 
\begin{figure}[!t]
\centering
 \includegraphics[width=0.8\linewidth]{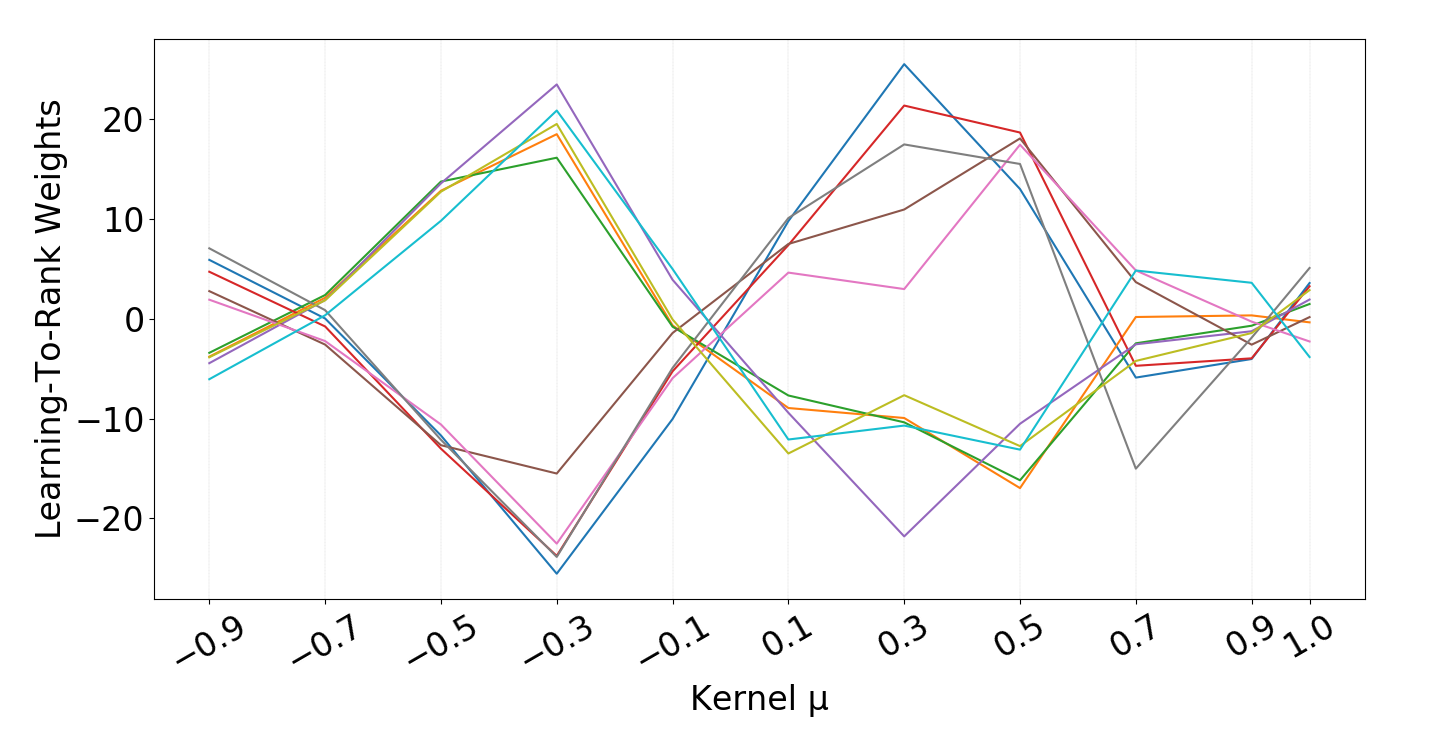}
  \caption{
  Learning to rank weights from 10 \texttt{K-NRM} trials.
  The X axis is the $\mu$ of a kernel. The Y axis is its ranking weight.
  \label{fig:ltr_weights}
}
\end{figure}

\begin{figure*}[th]
\centering
  \begin{subfigure}[b]{0.20\linewidth}
   
 \includegraphics[width=\linewidth]{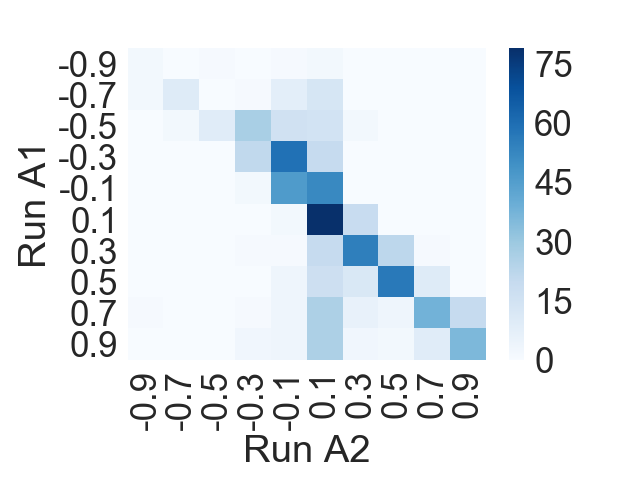}
    \caption{Run A1 vs. Run A2}
    \label{fig:heatmap_run23}
   \end{subfigure}
     \begin{subfigure}[b]{0.20\linewidth}
    \includegraphics[width=\linewidth]{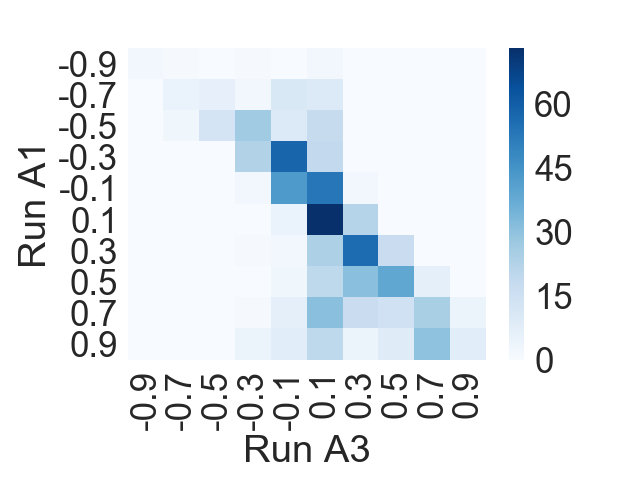}
    \caption{Run A1 vs. Run A3}
    \label{fig:heatmap_run25}
  \end{subfigure}
    \begin{subfigure}[b]{0.20\linewidth}
   \includegraphics[width=\linewidth]{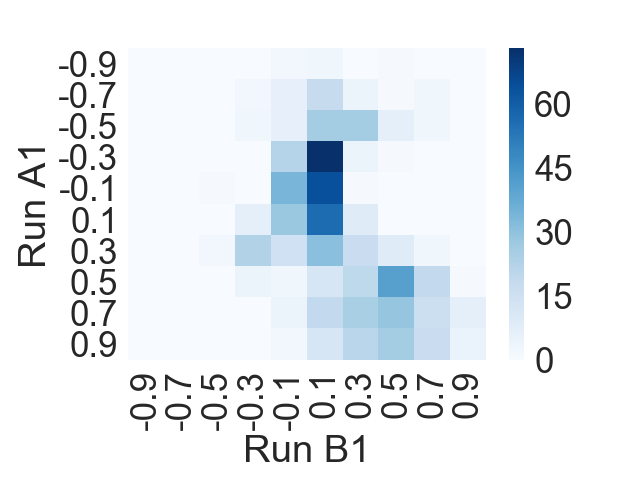}
    \caption{Run A1 vs. Run B1}
    \label{fig:heatmap_run21}
  \end{subfigure}
  \begin{subfigure}[b]{0.20\linewidth}
   
    \includegraphics[width=\linewidth]{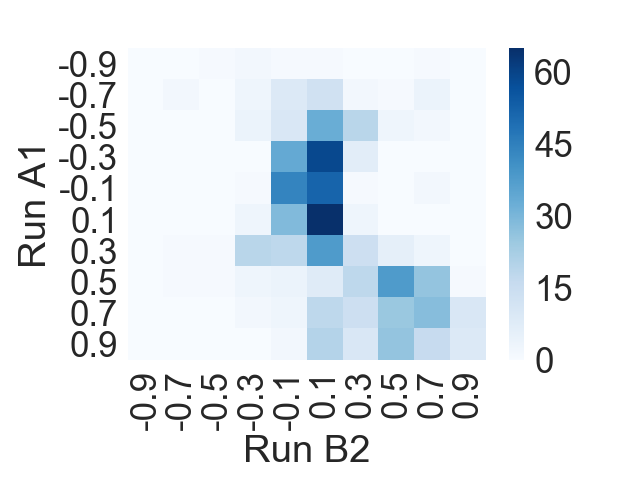}
    \caption{Run A1 vs. Run B2}
    \label{fig:heatmap_run24}
  \end{subfigure}

 \caption{
Word pair movements between runs from two patterns, A and B. 
% Pattern A includes Runs A1, A2, and A3. Pattern B includes B1 and B2. 
 Each cell $(\mu_x, \mu_y)$ in the heat map indicates the number of word pairs whose cosine similarities fall into Kernel $\mu_y$ in Run A1 (Y-axis) and  kernel $\mu_x$ in the other run (X-axis). 
%  Darker cell indicates more word pairs.
}
 \label{fig:heatmaps}
\end{figure*}

\begin{table*}
\centering
\caption{
The performance of ensemble models using different methods for selecting base models. \texttt{K-NRM} Mean is the average performance of 50 base models. $*, \dagger, \mathsection$ indicate statistically significant improvements (p < 0.05) over \texttt{K-NRM} Mean, \texttt{Ensemble-A} and \texttt{Ensemble-B} respectively. %Statistical significance
%was tested using the permutation test with p < 0.05. 
\label{table:ensemble-diff}
}
\setlength{\tabcolsep}{2pt} 
\begin{tabular}{@{}l|l|l|l||l|l|l||l@{}}
\hline
 & \multicolumn{3}{c||}{\textbf{Testing-SAME}} & \multicolumn{3}{c||}{\textbf{Testing-DIFF}} & \textbf{Testing-RAW} \\ \hline
\multicolumn{1}{c|}{\textbf{Model}} & \multicolumn{1}{c|}{\textbf{NDCG@1}} & \multicolumn{1}{c|}{\textbf{NDCG@3}} & \multicolumn{1}{c||}{\textbf{NDCG@10}} & \multicolumn{1}{c|}{\textbf{NDCG@1}} & \multicolumn{1}{c|}{\textbf{NDCG@3}} & \multicolumn{1}{c||}{\textbf{NDCG@10}} & \multicolumn{1}{c}{\textbf{MRR}} \\ \hline
\texttt{K-NRM} Mean & 0.2859  & 0.3495  & 0.4396 & 0.3242 & 0.3365  & 0.4378  & 0.3547 \\
Ensemble-A & 0.3270 (14\%)$*$ & 0.3824 (9\%)$*$ & 0.4691 (7\%)$*$ & 0.3702 (14\%)$*$ & 0.3694 (10\%)$*$ & 0.4573 (4\%)$*$ & 0.3908 (10\%)$*$ \\
Ensemble-B & 0.3215 (12\%)$*$ & 0.3723 (7\%)$*$ & 0.4570 (4\%)$*$ & 0.3831 (18\%)$*$ & 0.3749 (11\%)$*$ & 0.4629 (6\%)$*$ & 0.3930 (11\%)$*$ \\
Ensemble-A\&B & 0.3359 (17\%)\textsuperscript{$*\dagger \mathsection$ } &	0.3811 (9\%)\textsuperscript{$*\dagger \mathsection$} &	0.4689 (7\%)\textsuperscript{$*\dagger \mathsection$} & 	0.3931 (21\%)\textsuperscript{$*\dagger \mathsection$} &	0.3841 (14\%)\textsuperscript{$*\dagger \mathsection$} & 	0.4684 (7\%)\textsuperscript{$*\dagger \mathsection$} & 	0.4035 (14\%)\textsuperscript{$*\dagger \mathsection$}\\
%Ensemble All & 0.3303 (15\%) & 0.3826 (9\%) & 0.4669 (6\%) & 0.3950 (22\%) & 0.3866 (15\%) & 0.4699 (7\%) & 0.4045 (14\%) \\
\hline
\end{tabular}
\end{table*}

\section{Variance}
\label{subsection:multiple}

The first experiment studied the consistency of \texttt{K-NRM} by running 50 stochastically trained models with random initialization. The consistency among the 50 trials is examined at the query-set level and the individual query level. 

%The performance of the 50 models on the three metrics is summarized in Table \ref{table:stability-runs}. 
%The min/max differences can be large, especially for NDCG@1. However the standard deviations are small, ranging in 0.5-1.3\% absolute, and 1-4\% relative to mean values. We identified that the min/max differences are due to a small number of outliers and performance is stable across most trials. We also show the results reported by Xiong et al~\cite{xiong2017end} in Table 1. Their model performance falls into the lower end of our trials, probably due to different vocabularies and stopping conditions. 

The performance of 50 models on three metrics is summarized in Table \ref{table:stability-runs}. 
The min/max differences are large, especially for NDCG@1. However the standard deviations are small, ranging from 0.5-1.3\% absolute, and 1-4\% relative to mean values. The min/max differences are due to a small number of outliers.  Performance is stable across most trials. Table 1 also shows results reported by Xiong et al~\cite{xiong2017end}. Their model performance falls in the lower end of our trials, probably due to different vocabularies and stopping conditions.

The next analysis studied the consistency at the individual query level by examining   document rankings generated by different trials. For each query we examined the top $k$ ranked document from 10 different trials. The total number of distinct documents indicates how well the models agree about which documents to place at the top of the ranking. A histogram (Figure \ref{fig:agreement}) shows the number of queries at each agreement level for top 1, 3, and 10 ranked documents. 

Different trials rank different documents at the top to some extent. For about 50 of the 1000 queries, all 10 trials select the same document at rank 1 (Figure \ref{fig:histogram1}); for 35\% of the queries, the trials select 2-3 different documents. Moderate consistency is observed across the trials. Only 15\% of the queries get more than $5$ different documents from the 10 trials. None of the queries get 10 completely different documents at the top 1. %, which means that for every query at least 2 trials agree with each other on the most relevant document. 

% A similar trend is seen in Figure \ref{fig:histogram3}, where for 66\% of the queries, the 10 trials collectively select 3-9 different documents to fill the top 3 slots.  The document sets from the 10 trials converge deeper in the rankings. In the top 10 rankings (Figure \ref{fig:histogram10}), the histogram shifts to the left,  indicating that the 10 trials have higher agreement. This is expected because \texttt{K-NRM} only re-ranks the top 30 documents. The disagreement in the top 1 and 3 ranks indicates that even though different trials largely have the same sets of documents, their rankings are slightly different. 

Figure \ref{fig:histogram3} shows a similar trend. For 66\% of the queries, the 10 trials collectively select 3-9 different documents for the top 3 slots.  The document sets from the 10 trials converge deeper in the rankings. In the top 10 slots (Figure \ref{fig:histogram10}), the histogram shifts left,  indicating that the 10 trials have higher agreement. This is expected because \texttt{K-NRM} re-ranks the top 30 documents. The disagreement in the top 1 and 3 ranks indicates that although different trials have similar sets of documents, their rankings are slightly different. 
\section{Latent Matching Patterns} \label{subsection:variance}
To better understand the model differences, we investigated the model parameters through multiple \texttt{K-NRM} trials. \texttt{K-NRM} has two trainable components: the word embedding layer and the learning-to-rank layer. The word embedding layer aligns query-document word pairs and assigns them to the closest kernels by their cosine similarity. The learning-to-rank layer learns the importance  of word pairs around each kernel. This analysis studied both parameters.

Figure \ref{fig:ltr_weights} plots the learning-to-rank weights from 10 random trials. The trials fall into two main patterns. One pattern starts with a downward slope and then moves upward while the second pattern goes the other way.  \texttt{K-NRM} allocates word pairs into kernels based on their contribution to relevance. Different learning-to-rank weights indicate different ways of allocating word pairs to kernels. 

We further studied the two patterns with word embeddings from multiple trials. We randomly picked 5 runs. Runs A1--A3 belonged to one learning-to-rank weight pattern (Pattern A); runs B1--B2 belonged to the other pattern (Pattern B). We compared the word pair distribution between pairs of runs through a heat map 
with each cell $(\mu_x, \mu_y)$ indicating the fraction of word pairs that fall into kernel $\mu_x$ in one run and kernel $\mu_y$ in the other run. Figure \ref{fig:heatmaps} shows the heat maps between Run A1 and the rest of runs.  

Runs from the same pattern have similar heat maps. As shown in Figure \ref{fig:heatmap_run23} and \ref{fig:heatmap_run25}, Runs A2 and A3 show a strong diagonal pattern, indicating that most of the word pairs are in the same kernel as in run A1. 
Runs from pattern B share another word pair distribution. As can be seen from Figure \ref{fig:heatmap_run21} and \ref{fig:heatmap_run24}, 
a lot of word pairs are assigned to a different kernel by runs B1/B2 as compared to the kernel assigned by run A1. 
The results reveal two distinct latent matching patterns. Trials from the same pattern have similar learning-to-rank weights and word embeddings. The two patterns differ largely in their word pair alignment. 

Although the two patterns align word embeddings differently, both are equally effective and produce similar accuracy (Table 1).
\section{Ensemble Model}
The different rankings and distinct patterns in multiple \texttt{K-NRM} trials provided possibilities to reduce risk and improve the model's generalization ability using ensemble models ~\cite{krogh1995neural}. The following experiments studied the effectiveness of ensemble models. 

\begin{figure}[t]
\centering
 \includegraphics[width=0.5\linewidth]{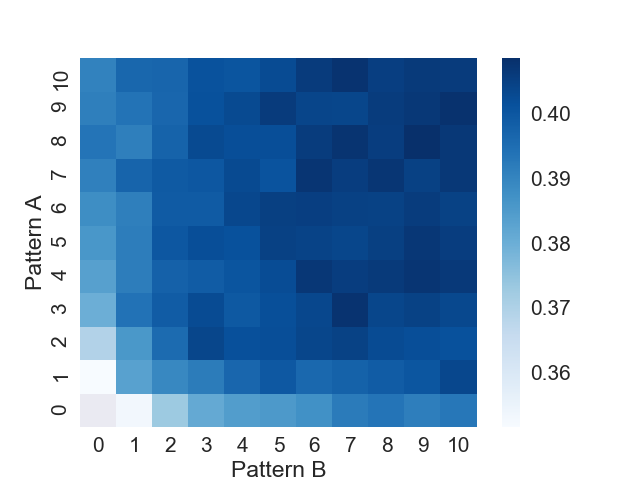}
  \caption{
  The accuracy of ensemble models that combine different numbers of base models from Patterns A and B.  Each cell is the MRR (Testing-RAW) of an ensemble model built with $m$  Pattern A models and $n$ Pattern B models.
%   Darker color indicates higher accuracy.
  \label{fig:ensemble_heatmap}
}
\end{figure}

We used an unweighted-average ensemble model~\cite{krizhevsky2012imagenet} that averages the scores from multiple trials. 
To investigate the effects of latent matching patterns, we tested different types of ensemble models: \texttt{Ensemble-A} used 10 base models randomly selected from Pattern A; \texttt{Ensemble-B} used 10 base models from Pattern B;  \texttt{Ensemble-A\&B} used base models from both patterns, 5 from each\footnote{We found that performance saturates with more than 10 base models.}.  To make evaluation reliable,   10 ensemble rankers were generated for each method with different base models randomly chosen from a pool of 50 \texttt{K-NRM} trials. 

All ensemble methods  significantly outperformed individual models (Table \ref{table:ensemble-diff}). The differences in document rankings allowed multiple trials to `vote' in the ensemble model. Documents favored by the majority of trials are voted up, whereas documents that are wrongly ranked high in a poor trial are voted down. 
Comparing NDCG scores at different depths, we see that ensembles are most effective at the top of the ranking. This is because the dataset mostly contains 20-30 documents per query. There is more opportunity for disagreement at the top, which gives more scope for improvement. The ensemble generalization ability is reflected by the improved performance on Testing-DIFF and Testing-RAW as compared to Testing-SAME (Table 2).  %As can be seen from in Table 2, the performance on the two test sets were both improved by large margins; Testing-DIFF gains more than Testing-SAME. 

\texttt{Ensemble-A\&B}  outperformed \texttt{Ensemble-A} and \texttt{Ensemble-B}  (Table 2), which indicates that having and recognizing two distinctive matching patterns is beneficial to ensemble models. 

To further understand the effects of the two patterns, we tested ensemble models with \texttt{m} Pattern A models and \texttt{n} Pattern B models. Figure \ref{fig:ensemble_heatmap} shows MRR on Testing-RAW as a heatmap. 
It confirms that having two variations enables better ensembles; ensemble models that only used one pattern have the lowest accuracy. Compared to single pattern ensembles, mixed ensembles can achieve the same accuracy using a smaller ensemble model with fewer base models. For example, cell $(3,3)$ has higher accuracy than cell $(10,0)$. Besides, ensemble models benefit from a balanced mix of the two patterns, as seen from the darker cells around the diagonal which have similar number of base trials from each pattern. 

Prior research did not recognize that \texttt{K-NRM} consistently converges to a small number of distinct, equally-good local optima.  Recognizing this helps in constructing high-quality ensembles.  
\section{Conclusion}
This paper studies the consistency and variation of \texttt{K-NRM}, a recent state-of-the-art neural ranking model. Unlike feature-based methods where features are stable and ranking models are often convex, neural networks are non-convex and employ stochastic training, making it important to consider the ranking stability in neural IR. By investigating multiple trials of \texttt{K-NRM}, we find that its accuracy is quite stable (has low standard deviation) in spite of its random components.  However, stable NDCG does not imply identical rankings at the individual query level.  Different trials have moderate agreement about which document to rank first.  
Ten trials collectively select 1-3 documents to rank first for 40\% of our queries.

Our analyses further demonstrate that multiple trials of \texttt{K-NRM} converge to two latent patterns that are about equally effective. Runs within the same pattern converge to similar ranking weights and word embeddings. This behavior was not recognized by prior work, and is worth additional study.

The distinct but equally effective matching patterns makes \texttt{K-NRM} a good fit for ensemble models. Recognizing different convergence patterns and selecting ensemble components equally from each pattern further improves \texttt{K-NRM}'s accuracy and ability to generalize. 
% Better understanding the convergence and divergence of the neural ranking model guilds the development of our ensemble approach, which significantly improves the effectiveness, stability, and generalization ability of K-NRM. 

%\todo{use shorten conference names. e.g. http://www.cs.cmu.edu/~cx/papers/JointSem.pdf}

\section{Acknowledgments}
This research was supported by National Science Foundation (NSF) grant IIS-1422676.  Any opinions, findings, and conclusions are
the authors' and do not necessarily reflect those of the sponsor.

\bibliographystyle{ACM-Reference-Format}
\bibliography{bibliography}

\end{document}